\documentclass{iaus}
\usepackage{graphics}

  \checkfont{eurm10}
  \iffontfound
    \IfFileExists{upmath.sty}
      {\typeout{^^JFound AMS Euler Roman fonts on the system,
                   using the 'upmath' package.^^J}%
       \usepackage{upmath}}
      {\typeout{^^JFound AMS Euler Roman fonts on the system, but you
                   dont seem to have the}%
       \typeout{'upmath' package installed. iaus.cls can take advantage
                 of these fonts,^^Jif you use 'upmath' package.^^J}%
      }
  \else
  \fi

  \checkfont{msam10}
  \iffontfound
    \IfFileExists{amssymb.sty}
      {\typeout{^^JFound AMS Symbol fonts on the system, using the
                'amssymb' package.^^J}%
       \usepackage{amssymb}%
       \let\le=\leqslant  
         
      }{}
  \fi

  \IfFileExists{amsbsy.sty}
    {\typeout{^^JFound the 'amsbsy' package on the system, using it.^^J}%
     \usepackage{amsbsy}}
    {\providecommand\boldsymbol[1]{\mbox{\boldmath $##1$}}}

\newsavebox{\astrutbox}
\sbox{\astrutbox}{\rule[-5pt]{0pt}{20pt}}

\newcommand{\cl}{Cl\,0413--65}
\newcommand{\ms}{MS\,1008--12}
\newcommand{\cldrei}{Cl\,0303$+$17}

\title[Outskirts of Galaxy Clusters: intense life in the suburbs]
      {Internal kinematics of spiral galaxies in distant rich galaxy clusters}

\author[K. J\"ager {\it et al.\/}]%
{K. J\"ager$^1$, A. B\"ohm$^1$, B.L. Ziegler$^1$,
J. Heidt$^2$, C. M\"ollenhoff$^2$\break
}

\affiliation{$^1$Universit\"ats-Sternwarte G\"ottingen, Germany,
email: jaeger@uni-sw.gwdg.de\\[\affilskip]
$^2$Landessternwarte Heidelberg, Germany}

\pubyear{2004}
\volume{195}
\pagerange{1--8}
\date{?? and in revised form ??}
\jname{Outskirts of Galaxy Clusters: intense life in the suburbs}
\editors{A. Diaferio, ed.}
\begin{document}

\maketitle

\begin{abstract}
We present our project on galaxy evolution in the environment 
of distant rich clusters aiming at disentangling the importance of specific 
interaction and galaxy
transformation processes from the hierarchical evolution of field galaxies. 
Spatially resolved MOS spectra were gained with VLT/FORS to analyze the 
internal kinematics of disk galaxies. First results are shown for the 
clusters MS 1008.1$-$1224 (z$=$0.30), Cl 0303$+$1706 (z$=$0.42), and 
Cl 0413$-$6559 (z$=$0.51). Out of 35 late type cluster members, 
13 galaxies exhibit a rotation curve of the universal form rising in the inner 
region and passing over into a flat part. The other members have peculiar 
kinematics. The 13 cluster galaxies for which a maximum rotation velocity 
could be derived are distributed in the Tully-Fisher diagram very similar to 
field galaxies from the FORS Deep Field with corresponding redshifts. The same 
is true for seven non-cluster galaxies observed in the cluster fields. 
The TF-cluster spirals do not show any significant luminosity evolution as 
might be expected from certain clusterspecific phenomena. Contrary to that, 
the disturbed kinematics of the non--TF cluster spirals indicate 
ongoing or recent interaction processes.
\end{abstract}

\firstsection 
\section{Introduction}
Galaxies in clusters reside in a special environment which is 
characterized by high galaxy densities, the presence of a hot intracluster 
medium (ICM) and vast amounts of dark matter.
This environment strongly influences the evolution of the cluster 
members superposed on the field galaxy evolution that arises from the 
hierarchical growth of objects and the declining star formation rates over 
cosmic epochs. Besides tidal interactions (including merging) as known from 
the field, cluster galaxies are affected by clusterspecific phenomena 
related to the structure of the cluster (like harassment) or to the ICM 
(like ram-pressure stripping). These mechanisms can cause substantial 
distortions of the internal kinematics of disk galaxies leading to 
``rotation curves'' (RCs) that no longer follow the universal 
form (Persic et al. 1996). Rubin et al. (1999), for example, classified half 
of their sample galaxies observed in the Virgo cluster
as kinematically disturbed (asymmetric or even truncated curves). 
On the other hand, many cluster galaxies follow a tight 
Tully--Fisher relation (TFR) (e.g. Giovanelli et al. 1997).

An important idea presently discussed came from the finding that
local clusters are dominated by elliptical and lenticular galaxies 
while the distant ones show a high fraction of spiral and irregular 
galaxies (Dressler et al. 1997): 
field spirals might fall into clusters and experience a 
morphological transformation into S0 galaxies.
Related to this scenario it is not yet clear whether the halo of dark matter 
and, therefore, the total mass and the RC of a galaxy can also be affected by 
certain interaction phenomena. 

First results on the kinematics of galaxies in distant clusters were presented 
by Metevier (2004) with 10 TF--galaxies 
in the cluster Cl\,0024$+$1654 ($z=0.40$) and by Milvang--Jensen et al.~(2003) 
with 8 spirals in MS\,1054.4--0321 ($z=0.83$).
While the first study sees a larger scatter of their sample galaxies in the TF
diagram without evidence for an evolution of the zero point, 
the second one finds a trend towards
brighter luminosities (0.5--1 $B$ mag) with respect to field spirals of
comparable redshifts. 
It is argued that either truncation
of star formation or starbursts may cause an increased or decreased 
mass--to--light ratio, respectively.

\section{Project and observations}
\begin{figure}
\centerline{\resizebox{!}{14cm}{\includegraphics{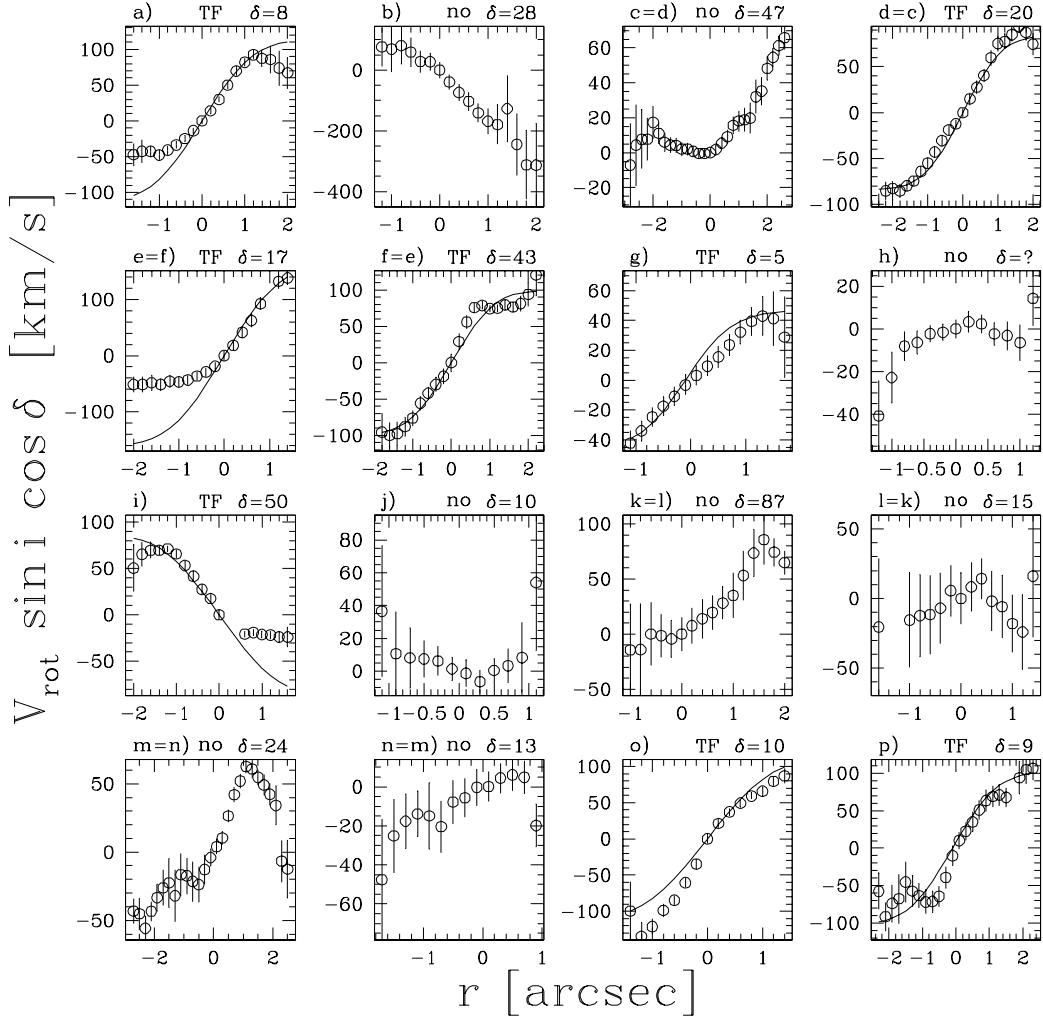}}}
  \caption{Spatially resolved velocity profiles for 12 disk 
galaxies in the cluster \ms\ at $z=0.30$. Note, panels c\&d, e\&f, k\&l 
and m\&n represent the same galaxy, rsp., but observed with different slit 
positions (the rotation angles of the two MOS masks differed by 67$^\circ$).
$\delta$ gives the angle between slit direction and apparent major axis.
Panels a--p are ordered according to the distance of the galaxies from the 
cluster center (upper left: smallest proj.~radius). 
Values of the maximum rotation velocity could be determined 
for cases a, d, e, f, g, i, o \& p indicated by the label ``TF''.
These galaxies enter the TFR shown in 
Fig.\,2. In all other cases, the kinematics are too disturbed.
In one case (m\&n), we observe a prominent bar that 
causes a highly asymmetric rotation curve.}
\end{figure}
\begin{figure}
\centerline{\resizebox{!}{12cm}{\includegraphics{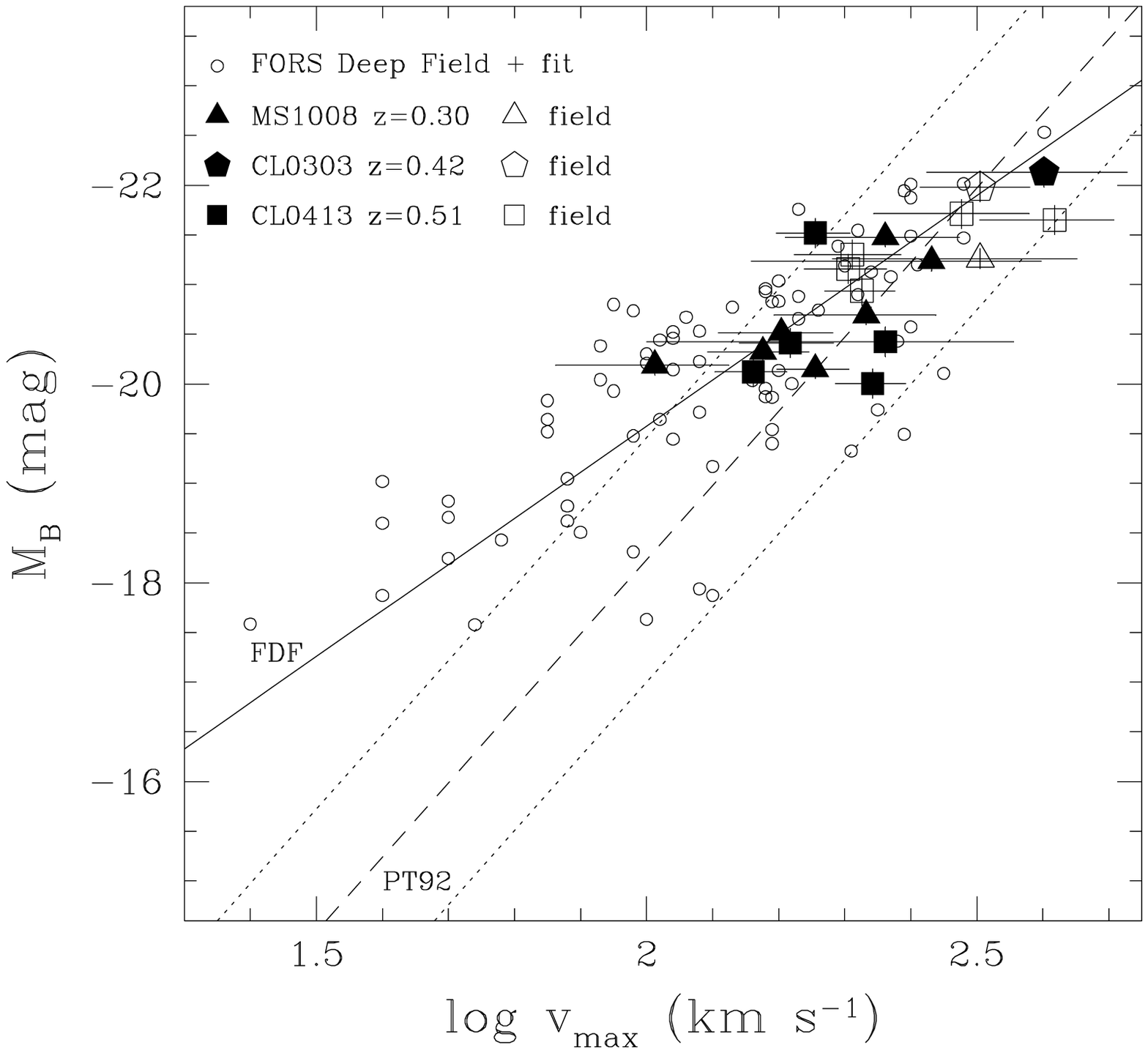}}}
  \caption{Tully--Fisher diagram of cluster spirals 
in \ms\ ($z=0.30$), \cl\ ($z=0.51$) and \cldrei\ ($z=0.42$). 
Also shown are 7 field
objects (open symbols) that were also observed in the clusters' 
field--of--view. In comparison to our FORS Deep Field sample of 77 field
galaxies with a mean redshift of 0.5 (small open circles), the cluster
galaxies are similarly distributed and do not deviate significantly from
the linear fit to the FDF sample (solid line).
The cluster members follow the same trend with respect to the local TFR
(the fit $\pm3\sigma$ to the Pierce \& Tully 1992 sample is given)
as the distant field galaxies. Since we observed mostly bright cluster
galaxies, their luminosities are not significantly increased in
accordance with an undisturbed evolution.
(Restframe $B$ magnitudes were calculated for a flat $\Omega_\lambda=0.7$ 
cosmology with $H_0=70$\,km\,s$^{-1}$\,Mpc$^{-1}$.)}
\end{figure}
During 5 nights at the ESO--VLT we gained spectra of 
galaxies within 7 clusters in the redshift range $0.3\le z<0.6$
with FORS\,1\&2. 
One of our goals was the derivation of spatially resolved RCs 
to analyse the internal kinematics of the galaxies
and to construct the TFR (Tully and Fisher 1977) 
of distant clusters.
HST/WFPC2 images of the core regions were retrieved from the archive 
to complement our own ground--based imaging. 
Here we present results of three of the clusters,
\cl\ ($z=0.51$) and \ms\ ($z=0.30$), \cldrei\ ($z=0.42$). 
Observations were made in 1999 and 2000
with FORS mounted at the Cassegrain focus of the VLT. 
We used grism 600R and get a spectral resolution of
$R\approx1200$ at a slitwidth of 1\,arcsec. The spatial scale 
was 0.2 arcsec/pixel with a dispersion of $\sim$1.08 \AA/pixel.
The total integration time was set to $\approx$2 hrs to meet our 
signal--to--noise requirements. Seeing conditions ranged between 0.4 and 1.3
arcsec FWHM. Two MOS setups have been observed
for each cluster, yielding 116 objects with apparent magnitudes
in the range $18.0<R<23$.
\section{Data reduction and rotation curve modelling} 
A detailed description of the sample selection and data reduction 
including finding charts and data tables can be found 
in J\"ager et al.~(2004).
For the derivation of position 
velocity diagrams 
we followed the way described in 
detail in B\"ohm et al. (2004) and Ziegler et al.(2003). 
Here we only give some short remarks on this
topic.

All RCs were determined using either the  
[O\,{\sc ii}]3727, H$\beta$ or [O\,{\sc iii}]5007 emission line. 
Gaussian fits have been applied stepping along the spatial axis
with a median filter window of typically 0.6 arcsecs to enhance the S/N. 
Since the apparent disk sizes of spirals at intermediate redshifts are
only slightly larger than the slit width (1\,arcsec), the spectroscopy
covers a substantial fraction of the two--dimensional
velocity field. Thus, the maximum rotation velocity $v_{\rm max}$ 
(the constant rotation in the outer part of a galaxy
due to the Dark Matter halo) cannot be determined ``straightforwardly'' from
the observed rotation.
To tackle this problem, we simulated the spectroscopy of each galaxies' 
velocity field with the respective inclination and position angle, also
taking seeing and luminosity weighting into account. 
The simulated rotation curves which best reproduced the
observed ones (solid lines in Fig.1) yielded the values of $v_{\rm max}$. 
\section{Results and discussion}
Redshifts and spectral types could be determined for 
89 galaxies. From the 50 
cluster members, 
35 turned out to be late type galaxies. 
Only 13 cluster galaxies exhibit a rotation
curve of the universal form rising in the inner region and 
passing over into a flat part. The other members have peculiar kinematics or 
too low S/N. For 7 non--cluster galaxies $v_{max}$ could also be measured.

Only those rotation curves which show no strong perturbations are eligible
for a determination of $v_{max}$
as needed
for the TF diagram which is shown in Fig.2. 
$v_{max}$ could be measured for 7(\ms), 5(\cl), and 
2(\cldrei) cluster members, respectively. 
Luminosities  were derived from total magnitudes (measured with SExtractor,
see Bertin \& Arnouts (1996)) 
of $V$ (\ms), $I$ (\cl), and $R$ (\cldrei) FORS images 
corrected for Galactic and intrinsic extinction, 
transformed to restframe Johnson $B$ according to SED type, and calculated for 
a flat $\Omega_\lambda=0.7$ cosmology ($H_0=70$\,km\,s$^{-1}$\,Mpc$^{-1}$).
 
Within our TF diagram the distant cluster spirals are distributed similar to
the FORS Deep Field (FDF) spirals (Ziegler et al. 2002, B\"ohm et al. 2004)
which have been observed with exactly the same instrument configuration and
which represent a similar cosmic epoch ($\langle z_{\rm FDF}\rangle=0.5$).
No significant deviation from the distant field TFR is visible and the cluster
sample has not an increased scatter, but the low number of cluster members 
prohibits any quantitative statistical analysis. 
We can conclude that
the mass--to--light ratios of the observed distant cluster spirals cover the
same range as the distant field population indicating that no clusterspecific
phenomenon dramatically changed the stellar populations. In particular, there
was no starburst in the recent past of the examined cluster galaxies that would
have significantly risen their luminosities. 
Since we mostly selected bright galaxies, the cluster members
occupy a region in the TF diagram where no significant luminosity evolution
is visible with respect to the local TFR.

On the other hand, this conclusion is true only for those objects that
enter the TF diagram and is not valid for the whole cluster sample. 
More than half of our cluster galaxies can not be used for a TF analysis 
due to their peculiar kinematics not following the ``rise--turnover--flat'' 
RC shape of large, isolated spirals. This is in contrast to the 
smaller fraction of peculiar curves in our sample of the FDF spirals.
The objects with peculiar RCs may actually be subject to ongoing or
recent interactions. 
Indeed, a morphological classification of the cluster 
spirals in terms of their asymmetry indices (Abraham et al.~1996) 
reveals that they have a slightly higher degree of asymmetry 
on the mean than 
the kinematically less disturbed ones. 
A correlation between morphological and kinematical disturbance
hints to a common origin for both in ``strong'' interactions like close encounters with
tidal effects, accretion events or even mergers. Such processes most probably
also influence the stellar populations of a galaxy changing its
integrated luminosity as well.

No trend is visible of the RC form with
(projected) clustercentric distance (cf.~Fig.1). 
But since all observed galaxies are 
located within the virial radius, this is in compliance with dynamical 
models in which the galaxy population of a cluster is well-mixed within that 
region. In particular, we most probably do not have any new arrivals from 
the field in our sample.\\ 

\begin{acknowledgments}
Based on observations with the ESO--VLT 
(64.O--0158 \& 64.O--0152). This work has been supported by the 
Volkswagen Foundation (I/76\,520) and the Deutsche Forschungsgemeinschaft 
(Fr 325/46--1 and SFB 439).
\end{acknowledgments}


\begin{thebibliography}{}

\bibitem[]{Abraham96}
     {Abraham et al.} 1996
     \textit{MNRAS} \textbf{279}, L47.

\bibitem[]{Bertin96}
     {Bertin E., \& Arnouts, S.} 1996
     \textit{A\&AS} \textbf{117},393.

\bibitem[]{Boehm}
{B\"ohm, A., Ziegler, B. L., Saglia, R.P., et al.} 2004
\textit{A\&A} in press (astro-ph/0309263)

\bibitem[]{Dressler}
{Dressler, A. et al.} 1997
\textit{ApJ} \textbf{490}, 577

\bibitem[]{Giovanelli}
{Giovanelli, R., Haynes, M. P., Herter, et al.} 1997
\textit{AJ} \textbf{113}, 22

\bibitem[]{Jaeger}
{J\"ager, K., Ziegler, B.L., B\"ohm, A., et al.} 2004
\textit{A\&A} submitted

\bibitem[]{Kodama}
{Kodama, T., \& Bower, R. G.} 2001
\textit{MNRAS} \textbf{321}, 18

\bibitem[]{Metevier}
{Metevier, A. J.} 2004. In Clusters of Galaxies: Probes of Cosmological
Structure and Galaxy Evolution, eds. Mulchaey, J.S, Dressler, A., Oemler, A. 
\textit{Carnegie Obs.~Astrophysics Series} \textbf{Vol.3}

\bibitem[]{Milvang}
{Milvang-Jensen, B., Arag\'on-Salamanca, A., Hau, G. K. T., et al.} 2003
\textit{MNRAS} \textbf{339}, L1

\bibitem[]{Persic}
{Persic, M., Salucci, P., \& Stel, F.} 1996
\textit{MNRAS} \textbf{281}, 27

\bibitem[]{Pierce}
{Pierce, M. J. \& Tully, R. B.} 1992
\textit{ApJ} \textbf{387}, 47

\bibitem[]{Rubin}
{Rubin, V. C., Waterman, A. H., Kenney, J. D. P.} 1999
\textit{AJ} \textbf{118}, 236

\bibitem[]{Tully}
{Tully, R.B., \& Fisher, J.R.} 1977
\textit{A\&A} \textbf{54}, 661

\bibitem[]{Ziegler}
{Ziegler, B.L., B\"ohm, A., Fricke, K.J., et al.} 2002
\textit{ApJL} \textbf{564}, L69 

\bibitem[]{Ziegler}
{Ziegler, B.L., B\"ohm, A., J\"ager, K., et al.} 2003
\textit{ApJL} \textbf{598}, L87 

\end{thebibliography}
\end{document}